# Search and design of nonmagnetic centrosymmetric layered crystals with large local spin polarization

Qihang Liu[1,2,*], Xiuwen Zhang[1], Hosub Jin[2], Kanber Lam[2], Jino Im[2], Arthur J. Freeman[2] and Alex Zunger[1,*]

[1]University of Colorado, Boulder, CO 80309, USA

[2]Deprtment of Physics and Astronomy, Northwestern University, Evanston, IL 60208, USA

[*]E-mail: qihang.liu85@gmail.com; alex.zunger@colorado.com.

**Abstract**

Until recently, spin-polarization in nonmagnetic materials was the exclusive territory of non- centrosymmetric structures. It was recently shown that a form of "hidden spin polarization" (named the "Rashba-2" or "R-2" effect) could exist in *globally* centrosymmetric crystals provided the individual layers belong to *polar point group* symmetries. This realization could considerably broaden the range of materials that might be considered for spin-polarization spintronic applications to include the hitherto 'forbidden spintronic compound' that belong to centrosymetric symmetries. Here we take the necessary steps to transition from such general, material-agnostic condensed matter theory arguments to material-specific "design principles" that could aid future laboratory search of R-2 materials. Specifically, we (i) classify different prototype *layered structures* that have been broadly studied in the literature in terms of their expected R-2 behavior, including the $Bi_2Se_3$-structure type (a prototype topological insulator), $MoS_2$-structure type (a prototype valleytronic compound) and $LaBiOS_2$-structure type (a host of superconductivity upon doping); (ii) formulate the properties that ideal R-2 compounds should have in terms of combination of their global unit cell symmetries with specific point group symmetries of their constituent "sectors"; (iii) use first-principles band theory to search for compounds from the prototype family of $LaOBiS_2$-type structures that satisfy these R-2 design metrics. We initially consider both stable and hypothetical $M'OMX_2$ ($M'$: Sc, Y, La, Ce, Pr, Nd, Al, Ga, In, Tl; $M$: P, As, Sb, Bi; $X$: S, Se, Te) compounds to establish an understanding of trends of R-2 with composition, and then indicate the predictions that are expected to be stable and synthesizable. We predict large spin splittings (up to $\sim 200$ meV for holes in $LaOBiTe_2$) as well as surface Rashba states. Experimental testing of such predictions is called for.

## I. Introduction: The local real-space view of hidden spin polarization in centrosymmetric nonmagnetic crystals

Traditionally, the origin of spin polarization in nonmagnetic solids has been attributed to spin-orbit coupling (SOC) within crystals that lack global inversion symmetry [1]. Crystals that possess instead global inversion symmetry would not have such a bulk Rashba-type (referred to as R-1), or Dresselhaus (referred to as D-1) polarization [2, 3] and would thus not be considered as candidates for spintronic applications. An alternative view has been proposed recently [4] that starts from the fact that since the SOC is a highly localized interaction anchored on atomic sites, it is the *local* environment (encoded in point group of sites) rather than the overall crystal symmetry that is an appropriate starting point for discussing SOC-induced spin polarization. Indeed, the inversion asymmetry in *local real space "sectors"* of a crystal leads to *local* spin polarization; the superposition of such polarizations from local sectors over the entire crystal would then produce the total (zero or nonzero) crystalline polarization. If the system lacks global inversion symmetry, the net effect would be nonzero, leading to the well-known bulk R-1 and D-1 effects with their respective distinctive spin textures. On the other hand, if the system as a whole does have a global inversion symmetry (centrosymmetric space groups) but the individual sectors are each inversion-asymmetric, still an interesting form of compensated polarization (see Fig. 1 a-h) could arise [4, 5] called R-2 and D-2, respectively

While the foregoing argument provided a material independent "proof of existence" for these effects, it did not tell the experimentalists where to look for such effects. The current work takes a step in the direction of aiding identification of such materials by systematically classifying the possible band structure and local spin polarization of different R-2 candidates, and then pointing to a group of materials that is predicted to have significant R-2 effect, showing thus where to look for such effects. Transitioning from material-independent to material-specific predictive theory could be helpful here because despite the importance of spintronics the material base currently used for it is rather narrow. This approach could broaden the type of materials that can be used and provide cross coupling between hidden spin polarization and other functionalities that also reflect spin-orbit physics (such as topological insulation or valleytronics) present in the same layered structures considered here for hidden spin polarization.

## II. Examples of prototypical groups of layered compounds that manifest R-2 and/or D-2 effects: $Bi_2Se_3$, $LaOBiS_2$ and $2H-MoS_2$

### A. Symmetry considerations

In Fig. 1a-h we consider the three leading classes of layered structures exemplified by the $Bi_2Se_3$-type (space group *R-3m*, Fig. 1f) structure, the idealized $LaOBiS_2$–type (*P4/nmm*, Fig. 1g) structure, and the $2H-MoS_2$-type transition-metal dichalcogenide (*P6$_3$/mmc*, Fig. 1h) structure. The unit cells of these structure types contain two, inversion asymmetric sectors that together form a centrosymmetric crystal. An intermediate buffer layer that contains an inversion center can separate these spin-orbit active inversion-partners. This intermediate section is supplied by the Se monolayer in $Bi_2Se_3$, the LaO monolayer in $LaOBiS_2$, or just vacuum in $2H-MoS_2$. Now, a single inversion asymmetric (IA) layer/sector α induces an effective magnetic field (owing to the existence of SOC in a low-symmetry environment [1]), and thus has spin polarization that is localized on that sector. If the crystal unit cell has an inversion center, there must be another sector β carrying local spin polarization with opposite spin texture. The corresponding energy bands must then be spin-degenerate due to the combination of inversion symmetry and time reversal symmetry, so the total spin, integrated over momentum ($\boldsymbol{k}$) space of these bands is zero. *But such global k-space compensation between the spins of the degenerate energy bands need not occur on a point-by-point basis in real space.* Indeed, explicit first-principles calculations of such a centrosymmetric crystal[4] have shown that the spin polarization is not locally compensated on each individual sector. Instead, there is distinct, residual spin on each sector. *This effect is not restricted to one unit cell but can exist in slab geometry, in which the surface states of both sides manifest opposite spin polarization.* This will be demonstrated later on in Sec IV and Fig. 6.

We refer to the hidden spin polarization in centrosymmetric bulk crystals made of sectors having a polar field as the *R-2 effect*[4], to be distinguished from bulk Rashba effect in non-centrosymmetric crystals (the *R-1 effect*) [8, 9]. Similarly, we refer to the hidden spin polarization in centrosymmetric bulk crystals made of IA sectors as the *D-2 effect,* to be distinguished from the D-1 effect in non-centrosymmetric crystals. In general, the crystal structure of R-2 material contains simultaneously polar field and inversion asymmetry, hence the coexistence of R-2 and D-2 effects[4]. Rather than being intrinsically absent, the R-2/D-2 effect in centrosymmetric crystals is concealed by global compensation, but present locally on a sector-by-sector basis. In particular, the R-2 effect is expected to be prevalent in *layered structures* because the IA sectors therein would experience a polar field along the stacking direction when there is nonequivalent bonding between the two sides of the spin-orbit active layer. In such structures one could induce the removal of the perfect compensation present in an R-2 crystal by applying a small electric field or creating a surface [6]. This R-2 effect broadens the range of bulk compounds that can be considered for such spintronic applications to include centrosymmetric crystals.

The general criteria for R-2 behavior are rather broad: centrosymmetric crystals with atomic sites that belong to polar group symmetry (point groups $C_1$, $C_2$, $C_3$, $C_4$, $C_6$, $C_{1v}$, $C_{2v}$, $C_{3v}$, $C_{4v}$, and $C_{6v}$). Such a broad definition applies to a large number of structure types include, for example the structures shown in Fig. 1f-h as well as to antiferroelectric materials in MgSrSi-type (*Pnma*) and NaNbO₃-type (*Pbcm*) structures [7]. Since these compounds have different characteristic crystal structures and band structures, it will be useful to develop more specific criteria for sorting out within these materials, which is likely to have the best R-2 characteristics. In this paper we will classify the different prototype band behavior of such R-2 layered materials, and then focus on the key ingredient that determines the quality of such R-2 materials and investigate the spin polarization features of a few compounds from the LaOBiS₂-type materials as favorable R-2 candidates.

**B. Classification of band structure prototypes for layered centrosymmetric R-2 crystals: Bi₂Se₃, LaOBiS₂ and 2H-MoS₂ types**

To get a better understanding of the band structure and spin localization based on different factors, we consider in Fig. 1 the evolution starting from one single IA sector that is a part of a centrosymmetric material.

In step (i) of Fig. 1, we consider a single sector α (e.g., a layer) obtained by cleaving the layered structure into two IA sectors. The isolated sector could be $(BiSe)^+$ layer in $Bi_2Se_3$, or the $(BiS_2)^-$ layer in $LaOBiS_2$, or the $MoS_2$ layer in $2H$-$MoS_2$. In the absence of SOC (Fig. 1a) each band is spin degenerate.

In step (ii) we introduce SOC into a single sector (Fig. 1b,c), which results in spin splitting due to the absence of inversion symmetry. We distinguish two types of wave vector $k$ and their vicinities: In Fig. 1b we consider $k$-points that are time reversal invariant (TRI), e.g., $\Gamma$ $(0, 0, 0)$ and $X$ $(0, 1/2, 0)$, where the band degeneracy at the crossing point between the two bands is protected by time reversal symmetry, whereas in Fig. 1c we show wave vector $k$ that is non-TRI e.g., $\boldsymbol{k} = K$ $(1/3, 1/3, 0)$ in the hexagonal Brillion zone of $2H$-$MoS_2$. Here the two spin-split bands do not cross but are nested within each other.

In step (iii) we add to sector α its inversion partner — sector β — having the same chemical composition but a reversed spatial configuration. Together they form the structure α + β that has global inversion symmetry (Fig. 1d through 1h). In Fig. 1d,e we assume that there is no interaction between sectors (e.g., infinitely separated sectors), so the ensuing band structure represents a superposition of the bands from both layers, leading to two-fold degenerate bands, each containing opposite spin polarizations localized on the respective individual sectors, as shown by red (α) and blue (β) arrows. Because α and β sectors do not communicate with each other, the local spin on each sector remains the same as if only that sector exists. Thus, in this ideal R-2 case we have distinct polarizations on each real-space sector. We next acknowledge in step (iv) that the sectors interact with each other, causing band repulsion and mixture of the local spin polarization between the sectors, thus diminishing relative to step (iii) the purity of the individual spin polarizations localized on each layer.

We next distinguish three material types here by discussing their R-2/D-2 effect [10] and the interplay between hidden spin polarization and their other functionalities:

**1. $Bi_2Se_3$: The case of strong coupling between individual R-1 layers leading to R-2 behavior**

Fig. 1f illustrates **$Bi_2Se_3$** (space group *R-3m*)**,** where the site point group of the two Bi atoms and the two side Se atoms belongs to the polar group $C_{3v}$, indicating an R-2 material; while that of the middle Se atom belongs to the centrosymmetric point group $D_{3d}$. We consider TRI *k* point Γ and *strong layer coupling* because of the proximity of these two (BiSe)$^+$ layers [11] and the absence of any effective intervening layer between them (an atomically -thin layer of Se). Here the coupling lifts the degeneracy, forming two isolated, non-crossing bands (unless the crossing point of two offsetting bands is protected by other symmetries such as fractional translation). Moreover, considering the $Bi_2S_3$ quintuple layer as a whole unit that consists two Rashba layer (BiSe)$^+$, such material has strong intra-unit coupling (between two (BiSe)$^+$ sectors) and weak inter-unit coupling (van der Waals interaction). Considering the band structure evolution from Fig. 1a to 1f we see that the inter-unit interaction is so strong that the upper band bends to form the conduction band with an inverted band curvature. A recent tight binding model showed that such layer-by-layer stacking of oppositly-aligned Rashba layers could lead to a bulk topological insulator, revealing the fundamental interplay between such R-2 structure and the topological physics [12].

Another consequence of the strong coupling between sectors is the spin mixture between different sectors, (shown in Fig. 1f) leading to a reduction in the net spin polarization localized on each sector. In other words, the formation of topological insulating behavior in $Bi_2Se_3$ requires strong coupling between spin-active sectors, which diminishes the residual spin on each sector.

## 2. 2H-$MoS_2$: The case of weak coupling between individual D-1 layers leading to D-2 behavior

Fig. 1h illustrates **2H-$MoS_2$** (space group *P6$_3$/mmc*), where we consider $\boldsymbol{k} = K$ which is non-TRI wavevector in the hexagonal structure. This structure has *weak layer coupling* because the two spin-active sectors ($MoS_2$ layer) are coupled by van der Waals interaction. For monolayer $MoS_2$, the site point group of Mo atom is $D_{3h}$ symmetry leading to D-1 behavior, while that of S atom is $C_{3v}$ symmetry leading to R-1 + D-1 behavior. However, the low-energy spectrum at the K valley of $MoS_2$-type material is dominated by the transition metal atom Mo, indicating that the spin physics in the vicinity of K is predominately a D-1 effect. Given that the spin component $S_z$ at non-TRI K valley is a good quantum number, such type of Dresselhaus spin splitting effect does not have a band crossing, but results in a Zeeman-like splitting in which the spins are out-of-plane and up and down spin locate at nested energy bands, respectively (see Fig. 1c) [13, 14].

Comparing with the D-1 case of monolayer of $MoS_2$, the bulk 2H-$MoS_2$ changes the space group of the unit cell but not that of the site point groups of Mo and S atom. Consequently, as shown in Fig. 2h, the spin polarization at K valley in 2H-$MoS_2$ is governed by the D-2 effect. The weak communication between inversion partners causes slightly larger separation of the nested bands and spin mixture from another layer, but most of the spin at the K valley remains on each layer.

The D-2 hidden spin polarization incorporated with the valleytronic physics in monolayer $MoS_2$ leads to many novel phenomena apparent in $n$ monolayers two-dimensional (2D) structures. For $n$ = odd the system is non-centrosymmetric and is expected to manifest effects related to this absence of inversion such as circular polarized luminescence originating from valley effects [15, 16]. In contrast, for $n$ = even (and for bulk 2H-$MoS_2$) inversion symmetry is present and thus no circular polarization is expected[15]. However, both theory [18] and experiments showed that circular polarized luminescence does exist in such centrosymmetric cases as an intrinsic feature due to the of the fundamental spin-orbit physics[17, 18]. This is because the significant D-2 effect causes spin-polarized absorption when excited by circular polarized light, and thus leads to imbalance luminescence for different helicity[17, 19].

### 3. LaOBiS$_2$: advantages and possible applications of the R-2 effect

Fig. 1g illustrates **LaOBiS$_2$** and related compounds. Here we discuss the properties of the $M'OMX_2$ compounds by assuming the nominal centrosymmetric structure *P4/nmm*. For this centrosymetric structure we consider TRI $k$ point $\boldsymbol{k}$ = X with *weaker layer coupling* because the two spin-active BiS$_2$ sectors in the unit cell are separated from each other by a barrier layer of La$_2$O$_2$. Here the bands still maintain a crossing at $k$ = X where interlayer coupling vanishes due to the fractional translation operation (non-symmorphic symmetry) that translates a layer to its inversion partner. In the vicinity of $X$, the band splitting is slightly enhanced by the layer interaction, comparing with that of step (iii), implying that the residual spin can mostly survive the interlayer coupling.

The discussion above indicates that case of LaOBiS$_2$ having weak interlayer coupling and strong SOC in each IA sector contains essentially two copies of Rashba spin polarization with significant but opposite spin textures localized on each sector, to which we will refer as favorable R-2 effect. This type of R-2 compensated spin polarization can have possible advantages over the bulk R-1 and the traditional two-dimensional (2D) Rashba effects regarding the manipulation of spin for potential applications. First, comparing with the ordinary 2D Rashba effect [3], the R-2 bulk-induced effect can have a larger spin splitting (more than 100 meV in LaOBiS$_2$) owing to the stronger polar field [9, 20] created by the intrinsic atomic configuration. Second, compared with the R-1 effect in non-centrosymmetric bulk materials such as BiTeI [8, 9], it is easier to manipulate the spin states of R-2 systems by selecting different spin-active sectors via small electric fields without reversing the internal polar field. Recently reported ferroelectric R-1 crystals such as GeTe could also manipulate the spin polarization due to the field-induced reversal of atomic displacements [20]. However, the corresponding switch speed in proposed spin-FET is much slower than that built on R-2 materials in which electric field controls the band structure and thus the spin state, because electron moves much faster than nucleus. Therefore, R-2 in LaOBiS$_2$ materials could offer advantages for the design of spintronic devices based on spin field effect transistors [21]. Third, the exploration of R-2 materials could significantly expand the range of materials with certain functionalities (e.g., second harmonic generation, photogalvanic effect, etc.) that is previously limited to low-symmetry non-centrosymmetric structures.

### III. Searching for other R-2 compounds within the $M$'O$MX_2$ family

We will next study the broad group denoted $M$'O$MX_2$ ($M$': group IIIa such as Sc, Y, La, Ce, Pr, Nd, or IIIb such as Al, Ga, In, Tl; $M$: group V such as P, As, Sb, Bi; $X$: group VI such as S, Se, Te). Our purpose is to identify the chemical and structural parameters that could control the quality of the R-2 effect, and then suggest that such bulk-induced spin polarization could be detected at the surface, indicating promising spintronic applications made of centrosymmetric crystals.

## A. Crystal structure

The nominal centrosymmetric structure *P4/nmm* ($T_0$) structure, which has equal Bi-$S_1$ bonds in *x-y* plane, was used in X-ray refinement of the crystal structure of the prototype material LaBiOS$_2$ first in 1995 [22], and in many other following experiments [23-25]. However, recently it was shown theoretically that phonon instability will make the two in-plane Bi-$S_1$ bonds in the *assumed P4/nmm* structure LaOBiS$_2$ non equivalent [26] so as to remove the instability. Such in plane bond inequivalence was verified by neutron diffraction experiment [27]. These observations exclude the *P4/nmm* structure as a possible ground state for LaBiOS$_2$ despite the tradition of fitting X-ray data to this model. We have considered the possible 3D stacking sequences of the in plane deformed 2D Bi-$S_1$ planes and found that three polytypes are possible: two non-centrosymmetric ($T_1$ of symmetry *P2₁mn* and $T_2$ of symmetry *C2*) and one centrosymmetric ($T_3$ of symmetry *P2₁m*). Upon total relaxation we find that for LaBiOS$_2$ the centrosymmetric structure $T_3$ is very slightly lower in energy than the other two structures. Thus, the final LaBiOS$_2$ structure is still either the ground state centrosymmetric $T_3 = P2₁m$ (so hidden R-2 spin polarization is expected) or a mixture of the polytypes that are non-centrosymmetric. In the latter case one can expect net spin polarization (D-1 or R-1) emerging from breaking the global inversion symmetry. We further examined some other LaOBiS$_2$-type materials and found that some of them have $T_0 = P4/nmm$ strcture as the ground state (e.g., LaOBiTe$_2$), while others have the $T_3 = P2₁m$ structure as the ground state (e.g., LaOSbS$_2$). All are thus centrosymmetric, albeit rather close energetically to the non-centrosymmetic polytypes. In the present study we assume uniformly the centrosymmetric $T_0 = P4/nmm$ structure for all *M'OMX$_2$* compounds so as to maintain straight comparison between different R-2 behaviors.

## B. Model Hamiltonian description of R-2 effect

Following the transition of band structure in Fig. 1, here we discuss the model Hamiltonian of single *MX$_2$* layer as a conventional Rashba system, and introduce the effect of inversion partner and interlayer coupling to get the band composition and the wave functions of the 2-fold degenerate band. Then we prove that there are two Rashba states with opposite spin texture localized on two $MX_2$ layers, respectively.

First, considering the relativistic Hamiltonian of an otherwise free electron in one $MX_2$ layer, with the SOC effect included

$$H = -\frac{\hbar^2}{2m}\nabla^2 + \frac{\hbar}{4m^2c^2}(\nabla V \times \boldsymbol{p}) \cdot \boldsymbol{\sigma} = -\frac{\hbar^2}{2m}\nabla^2 + \frac{\hbar^2\langle E_{in}\rangle}{4m^2c^2}(\sigma_x k_y - \sigma_y k_x) \quad (1)$$

where $m$, $c$, $V$, $\boldsymbol{\sigma}$ and $\langle E_{in}\rangle$ denote the electron mass, light speed, crystal potential, Pauli matrix vector ($\sigma_x, \sigma_y, \sigma_z$), and the effective built-in electric field (polar field induced by ionic bonding, see Fig. 2) along the $z$ direction, respectively. After solving Eq. (1), we get the energy splitting $E_\pm = \pm\alpha_R k$ caused by the relativistic term and the corresponding wavefunction $\psi_\pm = \frac{1}{\sqrt{2}}\begin{pmatrix} \pm e^{i\varphi} \\ 1 \end{pmatrix}$ leading to the helical spin fingerprint of Rashba splitting, where $\alpha_R = \frac{\hbar^2\langle E_{in}\rangle}{4m^2c^2}$ is the Rashba parameter, and $\varphi$ is the azimuth angle of electron momentum $\boldsymbol{k}$ in the $x$-$y$ plane.

Then we introduce a tight-binding Hamiltonian for the Rashba-bilayer system consisted of 4 $MX_2$ layers:

$$H_{BL} = \begin{pmatrix} R_+ & U & \mathbf{0} & T \\ U & R_- & T & \mathbf{0} \\ \mathbf{0} & T & R_+ & U \\ T & \mathbf{0} & U & R_- \end{pmatrix}. \quad (2)$$

Each term in the above matrix is a $2 \times 2$ matrix. In detail, $R_\pm$ present two $BiS_2$ layers that suffer opposite built-in electric fields:

$$R_\pm = \begin{pmatrix} \epsilon_{\boldsymbol{k}} & \pm\alpha_R(k_y - ik_x) \\ \pm\alpha_R(k_y + ik_x) & \epsilon_{\boldsymbol{k}} \end{pmatrix}, \quad (3)$$

where $\epsilon_{\boldsymbol{k}}$ denotes the on-site energy, $U = u_{\boldsymbol{k}}I_{2\times2}$ and $T = t_{\boldsymbol{k}}I_{2\times2}$ present the interaction between nearby $BiS_2$ layers though the $La_2O_2$ layer and the van der Waals layer, respectively. Given $M'_2O_2$ layer as a blocking barrier, the two adjacent $MX_2$ layers can be treated as an isolated bilayer electron system with van der Waals interaction between each other. Hence we ignore the interaction parameter between the two $MX_2$ layers through $M'_2O_2$ layer by assuming $u_{\boldsymbol{k}} = 0$. Therefore, regarding the global inversion symmetry of the system, in the representation of free electron in the $i$-th $MX_2$ layer $\psi_{\pm}^{(i)}$, the model Hamiltonian in Eq. (2) is simplified as

$$H_{BL} = \begin{pmatrix} \epsilon_{\boldsymbol{k}} & \alpha_R(k_y - ik_x) & t_{\boldsymbol{k}} & 0 \\ \alpha_R(k_y + ik_x) & \epsilon_{\boldsymbol{k}} & 0 & t_{\boldsymbol{k}} \\ t_{\boldsymbol{k}} & 0 & \epsilon_{\boldsymbol{k}} & -\alpha_R(k_y - ik_x) \\ 0 & t_{\boldsymbol{k}} & -\alpha_R(k_y + ik_x) & \epsilon_{\boldsymbol{k}} \end{pmatrix}. \quad (4)$$

By diagonalizing the matrix in Eq. (4), we get the two-fold degenerate eigenvalues

$$E'_{\pm}(\boldsymbol{k}) = \epsilon_{\boldsymbol{k}} \pm \sqrt{(\alpha_R k)^2 + t_{\boldsymbol{k}}^2}. \quad (5)$$

The corresponding wavefunction for $E'_{+}(\boldsymbol{k})$ as the form

$$\psi'_{+} = \begin{pmatrix} \sqrt{1-C_{\boldsymbol{k}}^2}\psi_{+}^{(1)} + C_{\boldsymbol{k}}\psi_{+}^{(2)} \\ C_{\boldsymbol{k}}\psi_{-}^{(1)} + \sqrt{1-C_{\boldsymbol{k}}^2}\psi_{-}^{(2)} \end{pmatrix}, \quad (6)$$

whereas the wavefunction for $E'_{-}(\boldsymbol{k})$ is written as

$$\psi'_{-} = \begin{pmatrix} -C_{\boldsymbol{k}}\psi_{+}^{(1)} + \sqrt{1-C_{\boldsymbol{k}}^2}\psi_{+}^{(2)} \\ \sqrt{1-C_{\boldsymbol{k}}^2}\psi_{-}^{(1)} - C_{\boldsymbol{k}}\psi_{-}^{(2)} \end{pmatrix}, \quad (7)$$

with

$$C_{\boldsymbol{k}} = t_{\boldsymbol{k}}[(\sqrt{\alpha_R^2 k^2 + t_{\boldsymbol{k}}^2} + \alpha_R k)^2 + t_{\boldsymbol{k}}^2]^{-\frac{1}{2}}. \quad (8)$$

The components of $\psi'_{+}$ and $\psi'_{-}$ represent the two degenerate states for the corresponding energy. From Eq. (5)-(8) we note that when $t_{\boldsymbol{k}}$ is small (van der Waals interaction), the two degenerate wavefunctions are dominated by the Rashba spin polarization localized on each $MX_2$ layer.

## C. First-principles calculation methods and electronic structure of $M'OMX_2$ compounds

All the first-principles calculations were performed with the Vienna *ab initio* package (VASP) [28]. The geometrical and electronic structures are calculated by the projector-augmented wave (PAW) pseudopotential [29] and the generalized gradient approximation of Perdew, Burke and Ernzerhof (PBE) to the exchange-correlation functional [30]. *Spin-orbit* coupling is all through calculated by a perturbation $\sum_{i,l,m} V_l^{SO} \vec{L} \cdot \vec{S} |l,m,i\rangle\langle l,m,i|$ to the pseudopotential, where $|l,m,i\rangle$ is the angular momentum eigenstate of $i$th atomic site [31]. The plane wave energy cutoff is set to 550 eV. Electronic energy minimization was performed with a tolerance of $10^{-4}$ eV, and all atomic positions were relaxed with a tolerance of $10^{-3}$ eV/Å. There are four variable cell-internal degrees of freedom, *i.e.*, the *z*-component of *M'*, *M*, $X_1$ (forming in-plane *M-X* bonds) and $X_2$ (forming perpendicular *M-X* bonds) site, that can be relaxed, while the other coordinates are protected by the crystal symmetry. In the bulk calculations, the lattice parameters are fully relaxed. In surface calculations, the atomic positions were relaxed with the lattice constant *a* fixed as the bulk one. The vacuum separation in the slab supercell is 20 Å to avoid the interaction between periodic images.

The triple-layer (TL) structure can be written as $(MX_2)^-/(M'_2O_2)^{2+}/(MX_2)^-$ made of two spin-active $MX_2$ rocksalt-like layers (sectors α and β) and an intermediate buffer layer $M'_2O_2$ separating them, with the fluorite-like structure (see Fig. 2a). The distance between two adjacent $MX_2$ layers is controlled by the van der Waals interaction. We show in Fig. 2c,d the band structures of bulk $M'OMX_2$ exemplified by the prototype compound LaOBiS$_2$ with its tetragonal Brillouin zone given in Fig. 2b. Here $MX_2 =$ BiS$_2$ and $M'_2O_2 =$ La$_2$O$_2$. From the calculated electronic structures we find that the low-energy spectrum originates mainly from the spin-active BiS$_2$ layers. Specifically, the six lowest conduction bands are dominated by Bi-*p* states, whereas the highest valence bands are composed of the S-*p* and Bi-*s* orbitals. On the other hand, the states originating from the La$_2$O$_2$ layer have little contribution to the bands near the Fermi level but appear at energies 3 eV above the conduction band maximum (CBM) or 1 eV below the valence band minimum (VBM) near the X valley.

*In the absence of SOC* (Fig. 2c), there is an extra two-fold band degeneracy along the symmetry line *X-M* and *R-A* for all the energy bands. This degeneracy results from the non-symmorphic aspect of the space group *P4/nmm*, i.e., from the participation of fractional translations (1/2, 1/2, 0) in the space group. Such fractional translation operation translates α (BiS$_2$) layer into β (BiS$_2$) layer along the real-space direction that corresponds to *X-M* direction in *k*-space, along which the interlayer coupling between chemically identical α and β BiS$_2$ layers vanishes.

*When turning on the SOC* (Fig. 2d), the band splitting along the Γ-*X* direction is caused by the combination effect of weak layer interaction (between two adjacent BiS$_2$ layers though van der Waals layer) and SOC-induced spin splitting, as shown in Fig. 2e and f. The layer interaction is dominantly from the channel through van der Waals layer, while the channel through $La_2O_2$ layer is nearly blocked. Therefore, The real-space separation of "different polarization on different sectors" would mostly survive interaction between the sectors that could mix the polarizations. Furthermore, the four-fold degeneracy along the *X-M* direction is reduced to two-fold, forming split bands that intersect at both the $k = X$ and $k = M$ wave vectors. This effect indicates the emergence of pure Rashba splitting on each of the two inversion partner layers, induced by the local IA environment felt by two $BiS_2$ sectors.

Using the tight-binding model discussed before we can easily understand the effect of SOC on the band structure of *M'OMX*$_2$. Without SOC, we have $t_{\boldsymbol{k}} = \alpha_R = 0$ along *X-M*, indicating four-fold degeneracy including spin from Eq. (5), while along $\Gamma$-*X*, $t_{\boldsymbol{k}} \neq 0$; thus, we have $C_{\boldsymbol{k}} = \frac{1}{\sqrt{2}}$, resulting in bonding and antibonding states according to Eq. (6) and (7). When SOC is available, we have $t_{\boldsymbol{k}} = 0$ but $\alpha_R \neq 0$ along *X-M*, which means $E'_{\pm}(\boldsymbol{k}) = \epsilon_{\boldsymbol{k}} \pm \alpha_R k$, $C_{\boldsymbol{k}} = 0$ and $\psi'_+ = \begin{pmatrix} \psi_+^{(1)} \\ \psi_-^{(2)} \end{pmatrix}$, $\psi'_- = \begin{pmatrix} \psi_+^{(2)} \\ \psi_-^{(1)} \end{pmatrix}$; thus, pure Rashba splitting emerges on each layer due to the local inversion asymmetry. We note that although the intervening $(La_2O_2)^{2+}$ layer does not contribute to the low-energy spectrum of eigenstates, it provides ionic charges that set up opposite polar fields pointing to α and β $(BiS_2)^-$ layers (along *z* and $-z$ direction, see Fig. 2), and thus make two $BiS_2$ layers form Rashba-type spin polarizations with opposite spin textures. Although the two copies of spin polarization with opposite spin states form degenerate bands in the *E vs. k* representation, each tends to be localized in real space on one of the two *MX*$_2$ layers.

### D. Choice of series of materials within the *M'OMX*$_2$ type

As noted above, a clear feature that could enhance the R-2 quality is the existence of an effective electronic separation between the two inversion partners, as illustrated in the transition from Fig 1d,e to Fig 1f-h. We will thus explore various barrier layers *M'*$_2$O$_2$ keeping its $2^+$ charge. Another important factor that determine the spin polarization is the magnitude of the SOC, anchored on different atoms in the spin-active $MX_2$ layer. We will thus consider different chemical identities for $M$ and $X$ and address how the magnitude of spin splitting depends on SOC of various atomic sites.

The O site has *non*-polar site point groups $S_4$ and thus does not contribute to the whole R-2 effect, so we do not change chemical identity for this site. Therefore, we classify the material groups into 3 series, using LaOBiS$_2$ as a prototype structure. In series I we vary the cation of the $MS_2$ layer in LaOMS$_2$ using $M$ = P, As, Sb, Bi. In series II we vary the anion of Bi$X_2$ layer in LaOBi$X_2$ using $X$ = S, Se, Te. In series III we vary chemical identity of the $M'_2O_2$ layer in $M'$OBiS$_2$ using $M'$ = Sc, Y, La, Ce, Pr, Nd, Al, Ga, In, Tl, Bi, and a recent reported compound SrFBiS$_2$ that has the same structure [32]. 7 of the 17 compounds studied here have been reported in the experimental literature to have the $P4/nmm$ structure[22, 32-34]. We include in our study 11 additional compounds (marked asterisks in Table I) that are unreported in the experimental literature but will be used to complete the theoretical trends in the computed R-2 features.

**E. Choice of metrics that reveal R-2 characteristics**

The Rashba parameter $\alpha_R$, defined by $\left| \frac{\partial E}{\partial k} \right|_{k=k_0}$, where $k_0$ is the wave vector of the band crossing point, is often used to assess the strength of the Rashba effect. However, for the functionality-driven materials screening used here, targeting large spin splitting, both the spin splitting energy $E_s$ and the corresponding momentum offset $k_s$, shown schematically in Fig. 2e,f and the insert of Fig. 4, are of interest. Note that we define both $E_s$ and $k_s$ along $X$-$M$ direction, because the band splitting along this direction results entirely from SOC. For R-2 materials without symmetry-protected band crossing, the band splitting consists both SOC-induced spin splitting and interlayer coupling. In this case $E_s$ could be defined as $\sqrt{E_S^2 - E_{NS}^2}$, where $E_S$ ($E_{NS}$) denotes the band splitting with (without) the implementation of SOC. Indeed, a larger $E_s$ is desirable for stabilizing and controlling spin while a larger momentum offset $k_s$ is advantageous for achieving a significant phase contrast between different spin and is thus favorable for the spin transport [21]. In contrast, $\alpha_R$ only evaluates the proportion of $E_s/2k_s$ for the ideal parabolic bands, but neither $E_s$ nor $k_s$ that are both useful for application.

**F. Predicted R-2 compounds from the *M'OMX*$_2$ family**

The band structures of the *M'OMX*$_2$ exhibit robust R-2 effects, as shown in Fig. 4. We list the band gaps ($E_g$) and main quantities related to R-2 splitting, including $E_s$, $k_s$ and $\alpha_R$, in Table I. Since the carrier types could be tuned by intrinsic defects or carrier injection in reality, here we consider the local spin splitting for both hole and electron states. Compared with electrons, the spin splittings of holes are generally more significant. Therefore, we plot the design metrics $E_s$ and $k_s$ for holes ($E_{s-h}$ and $k_{s-h}$, respectively) of different *M'OMX*$_2$ members classified into three categories, as shown in Fig. 5. We note the following main features:

*(i) R-2 characteristics in compounds with different spin-active MX$_2$ components*: The R-2 matrics of series II and I differ significantly through the different choices of the spin-active *MX*$_2$ layers. Specifically, Except LaOBiSe$_2$, when increasing the atomic number of *M* or *X* elements, all the quantities related to spin splitting, i.e., $E_{s-h}$, $k_{s-h}$ and $\alpha_{R-h}$, enhance with a broad range from LaOPS$_2$ to LaOBiTe$_2$ 0.3-196 meV, 0.001-0.073 Å$^{-1}$, and 0.07-2.43, respectively. Such spread implied that besides the local inversion asymmetric environment, the SOC on various atomic sites of the spin-active layer *MX*$_2$ is also a determining factor to the magnitude of favorable R-2 effect. With the atomic number increasing, the spin splitting effects become more remarkable.

*(ii) Effect of the buffer layer on R-2 characteristics*: Although the intermediate layer affects very little the low energy spectrum, we can still find variation of spin splitting accompanied with varying buffer layers. To explain this unexpected effect, we calculate the site dipole field of *M'*, *M* and *X* sites for all the compounds. Assuming the calculated static Coulomb potential near site $\boldsymbol{R_i}$ is $V^i(\boldsymbol{r})$, and then we can expand $V^i(r)$ by:

$$V^i(\boldsymbol{r}) = 0 + \sum_{\alpha=x,y,z} r_\alpha V_\alpha^i(0) + \frac{1}{2}\sum_{\alpha=x,y,z}\sum_{\beta=x,y,z} r_\alpha r_\beta V_{\alpha\beta}^i(0) + \cdots \quad (1)$$

The fist-order term [$V_\alpha^i$] denotes dipole field vector (its norm is $\sqrt{(V_\alpha^i)^2}$) that coupled with atomic SOC and thus induces local spin polarization. We note that in series III, despite the valence state of each layer is unchanged, the site dipole field of Bi reduces when carrying less electronegative elements, as shown in Table II. This might be a reason that causes the whole $E_{s-h}$ and $k_{s-h}$ to reduce.

*(iii) Contrasting the Rashba parameter $\alpha_{R-h}$ with the alternative metrics of $E_{s-h}$ and $k_{s-h}$*: For ideal parabolic bands, we have $k_s = m\alpha_s/\hbar^2$ and $E_s = 2m\alpha_s^2/\hbar^2$, so $\alpha_R$ here is evaluated by $E_s/2k_s$. However, as illustrated in Table I, the relationship between Rashba parameter $\alpha_{R-h}$ and $E_{s-h}$ or $k_{s-h}$ are not monotonic since $E_{s-h}$ and $k_{s-h}$ in real materials are also related to band curvature (i.e. effective mass $m$). For example, according to $\alpha_{R-h}$ it seems that BiOBiS$_2$ ($\alpha_{R-h} = 1.55$) has a better R-2 functionality than TlOBiS$_2$ ($\alpha_{R-h} = 1.46$). On the other hand, $E_{s-h}$ of TlOBiS$_2$ is 46% larger than that of BiOBiS$_2$, while $k_{s-h}$ 133% larger, apparently indicating that TlOBiS$_2$ is a better R-2 candidate. Other examples include the contrast between LaOSbS$_2$ (series I) and LaOBiSe$_2$ (series II), InOBiS$_2$, TlOBiS$_2$ and BiOBiS$_2$ (series III). Consequently, larger $\alpha_R$ does not definitely point to more significant spin splitting and better R-2 performance. Instead, both of $E_s$ and $k_s$ need to be addressed.

## IV. How can the R-2 effect be observed: Surface signatures

R-2 effect in such centrosymmetric layered materials could be observed if α and β sector are not equivalently detected, since the spin polarizations of the two sectors at this time are not exactly compensated by the global inversion symmetry. Especially, by truncating the bulk into surface states, we can dominantly detect the spin-active layers ($MX_2$) at the surface, and thus separate the bulk-induced spin polarization that hidden in 3D materials. It is also desirable to eliminate the interaction between two $MX_2$ layers and acquire an ideal Rashba effect in the 2D electron gas. Taking LaOBiS$_2$ as an example, due to the van der Waals interaction between adjacent BiS$_2$ layers, both of the bulk-truncated sides are BiS$_2$-terminated. Therefore, there is only one isolated BiS$_2$ layer at each surface since the buffer La$_2$O$_2$ layer blocks the interlayer coupling. Figure 6a plots the band spectrum of the 9-TL slab with a vacuum thickness of 20 Å between periodic images. We found that the electron surface states locate far away from the Fermi level. On the other hand, the hole surface states emerge from the bulk with significant band splitting, and decay rapidly into the slab, as shown in Fig. 6b,c.

Moreover, since the top and bottom sides are global inversion partners, the two-fold degenerate surface states are localized at the top and bottom surfaces with opposite spin patterns, analogous to the behavior of surface Dirac fermions in the inversion-symmetric topological insulators [35, 36]. The hole spin splitting energy and the corresponding momentum offset are 114 meV and 0.025 Å$^{-1}$, respectively. These values are quite similar to their counterparts in bulk calculation, demonstrating that the spin splitting here is predominately induced from the bulk structure, which is the same mechanism of the Rashba effect in some non-centrosymmetric bulk R-1 materials [8, 9, 20]. Such effect could be very robust and easy to reproduce since no additional requirements, such as artificial doping, intercalation, interface, or electric control is needed. The experimental evidence of the surface Rashba behavior by the SR-ARPES measurements is also expected.

## V Discussion: the effects of interlayer coupling on the R-2 characteristics

We note that whether bulk compounds or artificial superlattice, in R-2 systems made of IA inversion partners, the interlayer coupling plays a crucial role in determining the properties and potential applications of the whole system. In general, the interactions between two IA sectors form a pair of bonding and antibonding states, and opens a gap at the band crossing point. This IA bilayer scenario could lead to a novel phase diagram of topological superconductivity [37]. If the intralayer interaction (through the buffer layer) is strong while the interlayer interaction is relatively weak, e.g. van der Waals-like, the outer branch of the original bands could even begin to reverse due to the electron hybridization, offering possibilities for topological insulators with a new mechanism [12]. On the other hand, in the R-2 system $M'OMX_2$, we observe a pair of nearly perfect Rashba spin polarization, implying the immunity of states hybridization.

Increasing the interlayer coupling, i.e., compressing the materials along $z$-axis would enhance the mixture between different spin states of different layer and thus reduce the local spin polarization. On the other hand, it will also enhance the polar field felt by each sector and thus the spin splitting. In LaOBiS$_2$-type materials, the $M'_2O_2$ layer effectively blocks the interaction between two $MX_2$ layers. As a result, the local spin polarization is not sensitive on small uniaxial strain along $z$-axis. However, the local polar field and spin splitting could be tuned more significantly by such strain. For example, if we impose 5% compressive strain on LaOBiS$_2$ $z$-axis, the spin splitting energy for holes $E_{s-h}$ will increase 14%. Therefore, one can explore variety of tunable spintronic properties by controlling the coupling between IA layers in such systems, and thus pave an accessible avenue to design new functional materials or heterostructures.

**Acknowledgement**

Q.L. is grateful to Dr. Sonny Rhim and Dr. Longhua Li from Northwestern University for illuminating discussions. The work at CU Boulder was supported by NSF-DMREF Grant (No. DMREF-13-34170). This work used the Extreme Science and Engineering Discovery Environment (XSEDE), which is supported by National Science Foundation grant number ACI-1053575.

Table I: Band gap ($E_g$), energy splitting ($E_{s-h}$ and $E_{s-e}$), corresponding momentum offset ($k_{s-h}$ and $k_{s-e}$) and Rashba parameter ($\alpha_{R-h}$ and $\alpha_{R-e}$) for both holes and electrons (indicated in the inset of Fig. 4).

| M'OMX$_2$ Compounds | $E_g$ (eV) | $E_{s-h}$ (meV) | $k_{s-h}$ ($10^{-3}$Å$^{-1}$) | $\alpha_{R-h}$ (eV Å) | $E_{s-e}$ (meV) | $k_{s-e}$ ($10^{-3}$Å$^{-1}$) | $\alpha_{R-e}$ (eV Å) |
|---|---|---|---|---|---|---|---|
| **Series I** | | | | | | | |
| LaOPS$_2$[*] | 0.20 | 0.3 | 1.0 | 0.07 | 5.3 | 17 | 0.56 |
| LaOAsS$_2$[*] | 0.46 | 3.6 | 4.9 | 0.33 | 0.5 | 3 | 0.07 |
| LaOSbS$_2$[*] | 0.19 | 49 | 14 | 1.76 | 31 | 17 | 0.98 |
| LaOBiS$_2$ | 0.47 | 116 | 36 | 1.84 | 41 | 17 | 1.24 |
| **Series II** | | | | | | | |
| LaOBiS$_2$ | 0.47 | 116 | 36 | 1.84 | 41 | 17 | 1.24 |
| LaOBiSe$_2$ | 0.32 | 106 | 34 | 1.72 | 8.0 | 7.3 | 0.52 |
| LaOBiTe$_2$[*] | 0.10 | 196 | 73 | 2.43 | 5.5 | 8.0 | 0.32 |
| **Series III** | | | | | | | |
| ScOBiS$_2$[*] | 0.48 | 9.9 | 9 | 0.51 | 68 | 25 | 1.37 |
| YOBiS$_2$[*] | 0.40 | 80 | 22 | 1.86 | 12 | 8.7 | 0.65 |
| LaOBiS$_2$ | 0.47 | 116 | 36 | 1.84 | 41 | 17 | 1.24 |
| CeOBiS$_2$ | 0.47 | 108 | 32 | 1.80 | 37 | 16 | 1.07 |
| PrOBiS$_2$ | 0.46 | 108 | 30 | 1.89 | 35 | 15 | 1.07 |
| NdOBiS$_2$ | 0.45 | 105 | 28 | 1.93 | 32 | 14 | 1.05 |
| AlOBiS$_2$[*] | 0.60 | 21 | 43 | 0.30 | 243 | 96 | 1.16 |
| GaOBiS$_2$[*] | Metallic | 26 | 19 | 0.67 | 27 | 19 | 0.68 |
| InOBiS$_2$[*] | 0.25 | 97 | 30 | 1.74 | 15 | 11 | 0.67 |
| TlOBiS$_2$[*] | Metallic | 105 | 42 | 1.46 | 31 | 17 | 0.92 |
| BiOBiS$_2$ | 0.53 | 72 | 18 | 1.55 | 12 | 9.6 | 0.60 |
| SrFBiS$_2$ | 0.54 | 117 | 36 | 1.85 | 41 | 17 | 1.23 |

*Materials not reported to exist in the experimental literature

Table II: Site dipole fields on $M$, $X_1$ (forming in-plane $M$-$X$ bonds), $X_2$ (forming perpendicular $M$-$X$ bonds) and $M'$ sites

| M'OMX$_2$ Compounds | $E_M$ (V/Å) | $E_{X1}$ (V/Å) | $E_{X2}$ (V/Å) | $E_{M'}$ (V/Å) |
|---|---|---|---|---|
| **Series I** | | | | |
| LaOPS$_2$[*] | 1.6 | 0.6 | -1.4 | -2.7 |
| LaOAsS$_2$[*] | 2.4 | 0.6 | -1.3 | -2.7 |
| LaOSbS$_2$[*] | 1.1 | 0.4 | -1.1 | -2.6 |
| LaOBiS$_2$ | 1.8 | 0.4 | -1.1 | -2.6 |
| **Series II** | | | | |
| LaOBiS$_2$ | 1.8 | 0.4 | -1.1 | -2.6 |
| LaOBiSe$_2$ | 1.4 | 0.8 | -1.3 | -2.2 |
| LaOBiTe$_2$[*] | 1.0 | 0.7 | -0.7 | -1.8 |
| **Series III** | | | | |
| ScOBiS$_2$[*] | 1.3 | 0.9 | -1.0 | -1.3 |
| YOBiS$_2$[*] | 1.7 | 0.7 | -1.0 | -1.9 |
| LaOBiS$_2$ | 1.8 | 0.4 | -1.1 | -2.6 |
| CeOBiS$_2$ | 1.8 | 0.4 | -1.0 | -0.5 |
| PrOBiS$_2$ | 1.8 | 0.5 | -1.0 | -0.8 |
| NdOBiS$_2$ | 1.8 | 0.5 | -1.0 | -0.5 |
| AlOBiS$_2$[*] | 0.8 | 1.2 | -0.8 | -1.3 |
| GaOBiS$_2$[*] | 1.3 | 1.0 | -1.2 | 0.1 |
| InOBiS$_2$[*] | 1.5 | 0.8 | -1.2 | -0.1 |
| TlOBiS$_2$[*] | 1.7 | 0.7 | -1.3 | -0.3 |
| BiOBiS$_2$ | 0.8 | 0.3 | -1.8 | -0.6 |
| SrFBiS$_2$ | 1.0 | 0.3 | -0.8 | -1.1 |

*Materials not reported to exist in the experimental literature

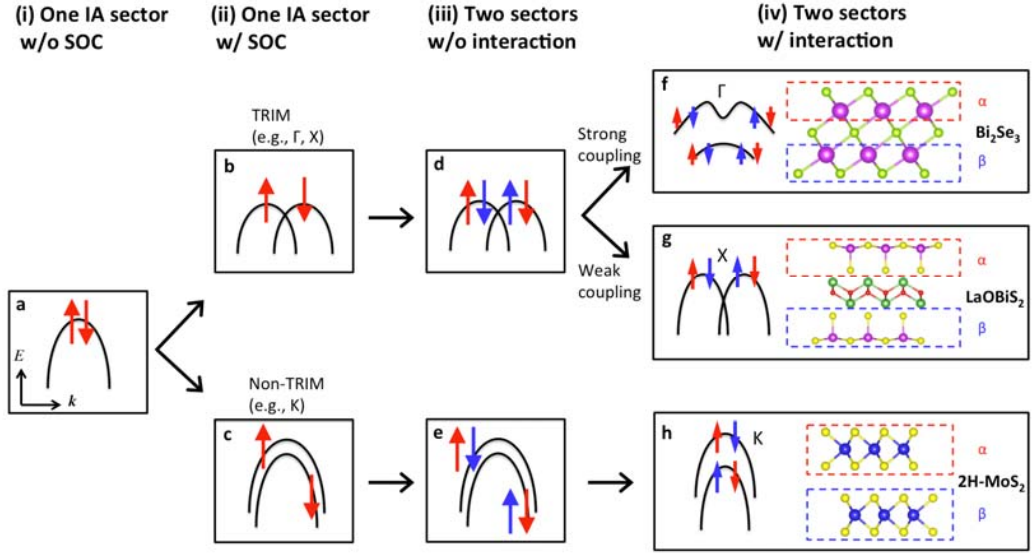

**Fig. 1:** Schematic, step by step depiction of the transition of the band structure and residual spin localized on each real-space sectors for different R-2 layer materials. There are four steps. (i) A single real-space sector α with inversion asymmetry IA. (ii) The influence of SOC combined with IA leads to two kinds of spin splitting, depending on different types of wave vectors. (iii) Here we add another IA sector β with the same composition as α forming inversion partner to sector α, assuming no interaction between layers α and β. (iv) Interlayer interaction leads to additional band splitting and the reduction of the projected spin polarization on each sector. The red and blue arrows denote the projected spin polarization on α and β sectors, respectively.

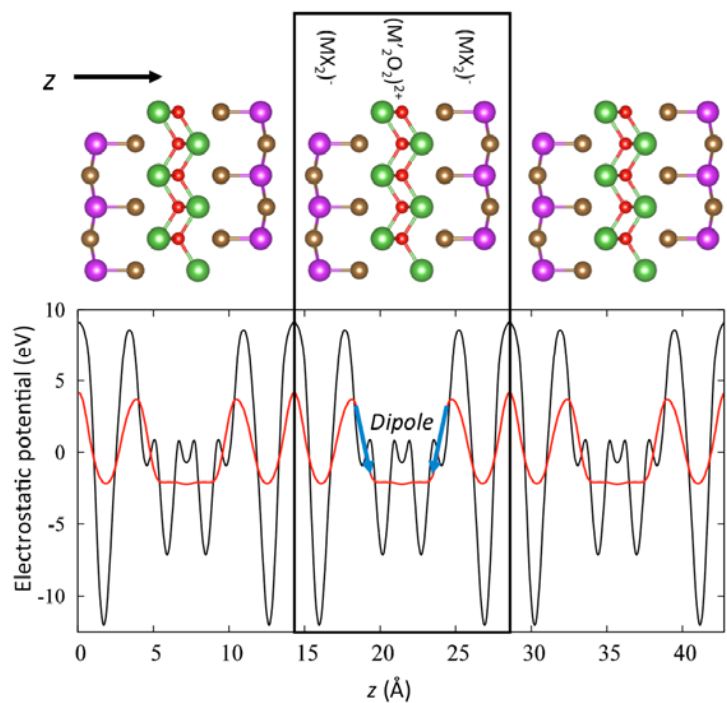

**Fig. 2:** Local electrostatic potential (ionic plus Hartree) integrated in $x$-$y$ plane (black solid) and its smooth profile by macroscopically averaging throughout a 3.8 Å-length (red solid). Blue arrows denote the opposite polar fields induced by ionic bonding between $(M'_2O_2)^{2+}$ buffer layer and two $(MX_2)^-$ spin-active layers.

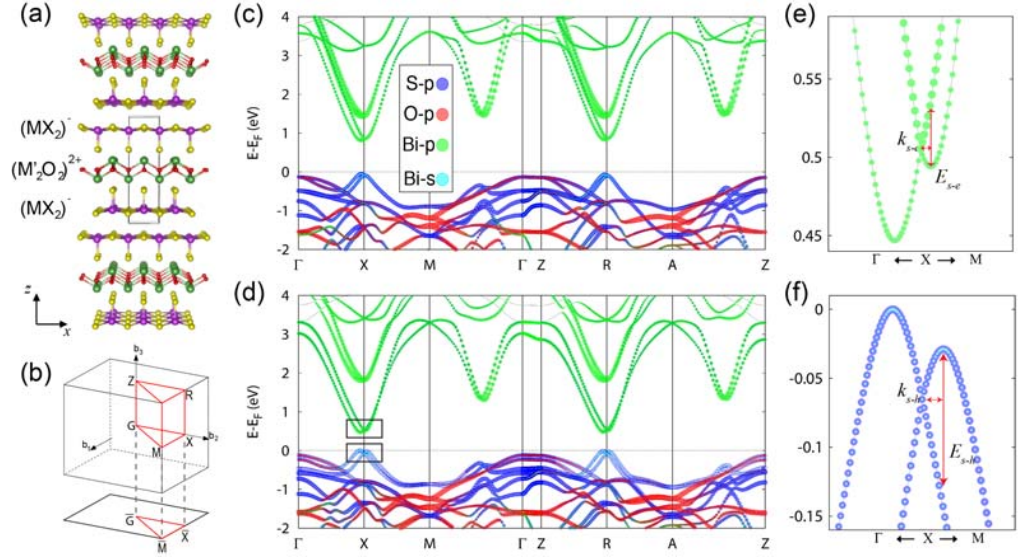

**Fig. 3:** (a) Crystal structure of $M'OMX_2$ family, with the triple-layer unit shown in the black frame. The green, red, purple, yellow balls represent $M'$, O, $M$, and $X$ atoms, respectively. (b) Brillouin zone of the tetragonal structure with the high-symmetry points projected onto the (001) surface. Band structures with atomic projection of a prototype material LaOBiS$_2$ (c) without SOC and (d) with SOC. The valence band maximum is set to zero. The scaled-up view of the band structure corresponding to the small frame marked in panel (d) at CBM (e) and VBM (f).

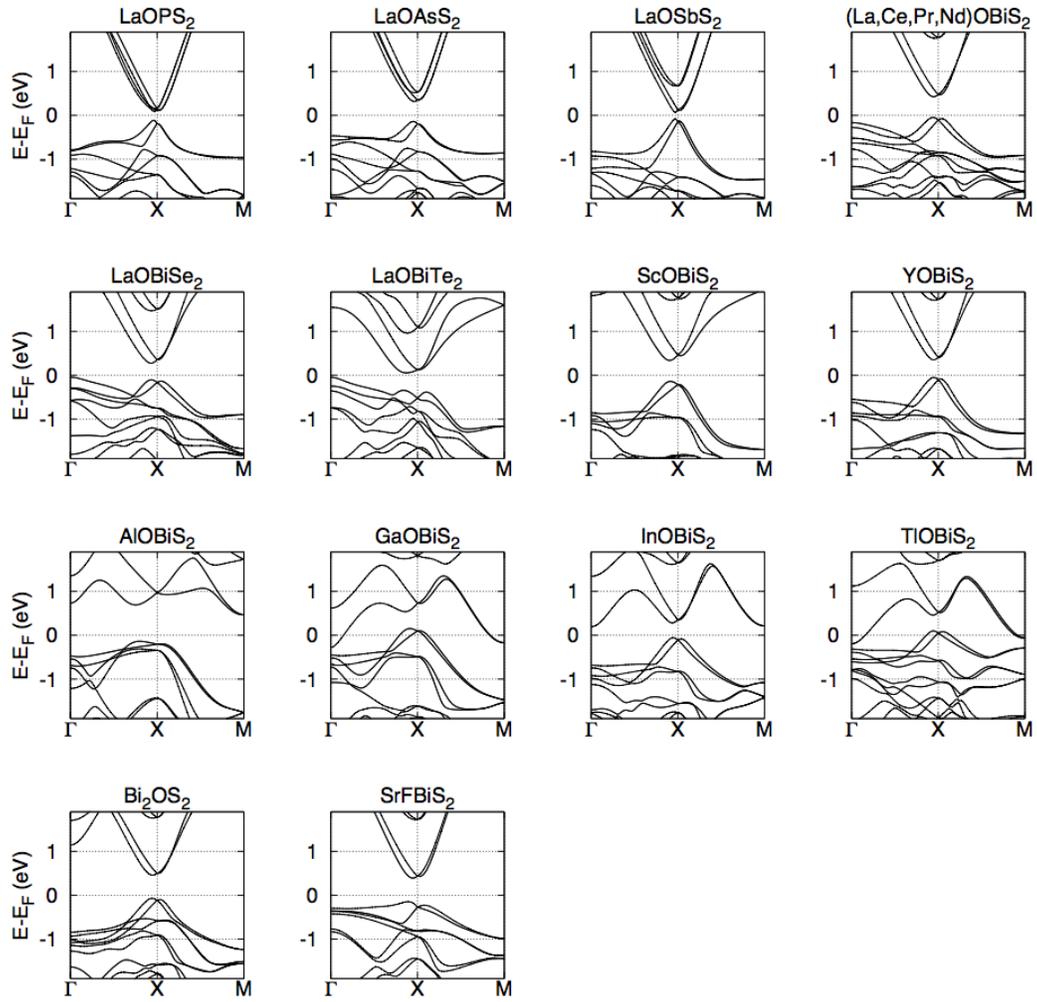

**Fig**. 4: Band structures of all the $M'OMX_2$ compounds. Note that the $M'OBiS_2$ (M' = La, Ce, Pr, Nd) compounds have very similar band structures, so we fold them into one panel.

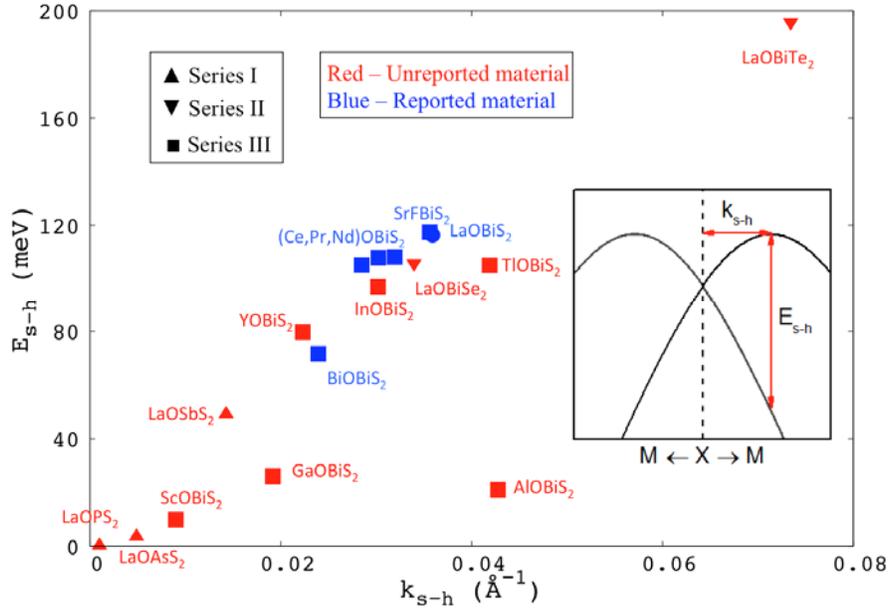

**Fig. 5:** Local spin splitting energy $E_{s\text{-}h}$ and corresponding momentum offset $k_{s\text{-}h}$ for holes (definition indicated in the inset) of 17 $M'OMX_2$ compounds. The blue and red symbols denote reported and unreported compounds under this structure, respectively.

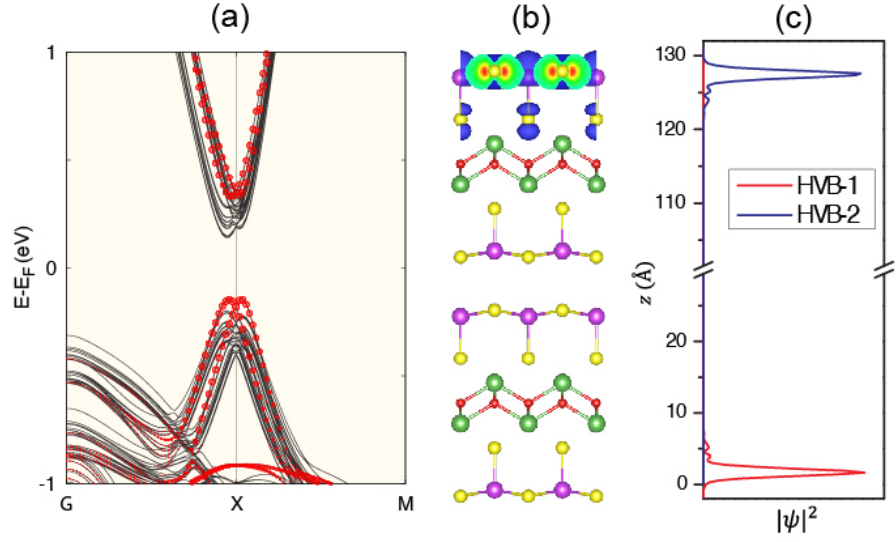

**Fig. 6:** (a) Band structure of the 9-TL LaOBiS$_2$ slab with SOC turned on. The red circles represent the surface states. (b) The charge density distribution of surface states, and (c) the corresponding real-space charge density integrated over (*x, y*) planes of the two-fold degenerate highest valence band (HVB-1 and -2).